\documentclass[iop]{emulateapj}
\usepackage{subfigure}
\makeatletter
\newcommand\ionpat[2]{#1$\;${\scshape{#2}}}

\def\halflight{10.9 kpc} 
\def\factoroffset{9}
\def\projoffset{60 kpc}
\def\massfaint{$7.9\pm0.7$}

\def\massfaintpow{10$^{7.9\pm0.7}$}
\def\massbrightpow{10$^{10.8\pm0.1}$}
\def\sfrfaint{0.05 M$_{\odot}$ yr$^{-1}$}
\def\metfaint{8.16}
\def\veloffset{60 km s$^{-1}$}
\def\veldisp{366}
\def\halphasone{$7514.4 \pm 0.3$ (statistical)}
\def\halphastwo{$7515.4 \pm 0.7$ (statistical)}

\submitted{Submitted to the Astrophysical Journal Letters}

\begin{document}

\title{Evidence that Gamma-ray Burst 130702A Exploded in a Dwarf Satellite of a Massive Galaxy}
\shorttitle{Host Galaxy of GRB 130702A}

\author{Patrick L. Kelly, Alexei V. Filippenko, Ori D. Fox, Weikang Zheng, and Kelsey I. Clubb}
\affil{Department of Astronomy, University of California, Berkeley, CA 94720-3411, USA}
\email{pkelly@astro.berkeley.edu}


\begin{abstract}
GRB 130702A is a nearby long-duration gamma-ray burst (LGRB) discovered by the \textit{Fermi} satellite whose associated afterglow was detected by the Palomar Transient Factory.  Subsequent photometric and spectroscopic 
monitoring has identified a coincident broad-lined Type Ic supernova (SN), and nebular emission detected near the 
explosion site is consistent with a redshift of $z = 0.145$.
The SN-GRB exploded at an offset of $\sim7.6''$ from the center of an inclined $r=18.1$ mag red disk-dominated galaxy, and $\sim0.6''$ from the center of a much fainter $r=23$ mag object. 
We obtained Keck-II DEIMOS spectra of the two objects and find a $2\sigma$ upper limit on their line-of-sight velocity offset of $\lesssim$\,\veloffset. 
If we calculate the inclination angle of the massive red galaxy from its axis ratio and assume that its light is dominated by a very thin disk,
the explosion would have a $\sim$\,\projoffset~central offset, or $\sim$\,\factoroffset~times the galaxy's half-light radius.
A significant bulge or a thicker disk would imply a higher inclination angle and greater central offset. 
The substantial offset suggests that the faint source is a separate, dwarf galaxy. 
The star-formation rate of the dwarf galaxy is $\sim$\,\sfrfaint, and we place an upper limit on 
its oxygen abundance of 12 + log(O/H) $<$ \metfaint~dex. 
The identification of an LGRB in a dwarf satellite of a massive, metal-rich primary galaxy suggests that recent detections of LGRBs spatially coincident with 
metal-rich galaxies may be, in some cases, superpositions. 

\end{abstract}

\maketitle 

\section{Introduction}

The high-energy photons and power-law afterglow characteristic of long-duration gamma-ray bursts (LGRBs), as well as the existence of 
coincident broad-lined Type Ic supernovae (SNe~Ic-BL; \citealt{ga98}; \citealt{ma03}; \citealt{st03}; \citealt{hj03}), are consistent with the creation of a relativistic jet after the core
of a massive star collapses to form a compact object \citep{wo93}. 
Theoretical modeling of possible progenitor systems strongly favors massive, 
rapidly spinning stars \citep{hirschimeynet05, yoonlanger05, woosleyheger06},
because high angular momentum is likely required to form the relativistic jets. 
These massive stars must also lose their outer hydrogen and helium envelopes, or burn them through internal mixing \citep{yoonlanger05, woosleyheger06}, because all associated SNe have had Type Ic-BL spectra which show
no evidence for hydrogen or helium. 


The metal content of LGRB host galaxies has received special attention because of the strong 
effect of metals on stellar mass loss rates. 
The outflowing stellar winds from massive stars grow in strength with increasing metallicity ($Z$) of
the envelope \citep{kudritzki00} approximately as $\dot{M} \propto Z^{0.66}$ \citep{vinkdekoter05}.
Single Wolf-Rayet stars with low metallicity are expected to have higher rates of rotation than their high-metallicity counterparts,
and a reasonable expectation might be that LGRBs would occur more frequently in metal-poor environments,
where single stars may retain more of their angular momentum before core collapse.
However, recent observations have suggested that many massive stars have binary companions (e.g., \citealt{sanademink12}).
Interaction with a close companion may be able to produce stars with the high angular momentum and stripped envelopes necessary for an LGRB \citep{zhangfryer01, fryerheger05}, even in comparatively metal-rich environments. 

\citet{mod08} found that SNe~Ic-BL with an LGRB strongly prefer more metal-poor hosts than those without a coincident LGRB (see also \citealt{stanek06}; \citealt{grahamfruchter12}).
Recent host-galaxy follow-up observations, however, have found examples of LGRBs that exploded at positions coincident with metal-rich host galaxies (e.g., \citealt{prochaska07}; \citealt{lev10highmet}; \citealt{savagliorau12}; see, among others, \citealt{kocevski11} for possible interpretations).


Here we show an example of an LGRB that probably exploded in a metal-poor dwarf satellite of a 
metal-rich galaxy. 
A plausible possibility is that a fraction of the LGRBs whose positions have coincided with metal-rich 
galaxies may have less luminous, metal-poor host galaxies that are difficult to 
detect. 
The low redshift ($z=0.145$) of GRB 130702A allows the separation of the dwarf neighbor host
from a massive primary galaxy in ground-based images, and 
makes possible the spectroscopic detection of weak emission lines from the dwarf galaxy's metal-poor gas. 
Section \ref{sec:detection} briefly discusses the discoveries and multi-wavelength observations of the GRB, coincident optical transient, and associated SN Ic-BL by several observing teams. 
Section \ref{sec:data} describes photometry and spectroscopy of the sources 
close to the SN-GRB site. 
We present results from our photometric and spectroscopic analyses in \S \ref{sec:results}, and we 
provide both a summary and interpretation in \S \ref{sec:discussion}.

\section{Detection and Follow-up Observations}
\label{sec:detection}
GRB 130702A was detected by the \textit{Fermi}-GBM \citep{gcn14972} and \textit{Fermi}-LAT
\citep{gcn14971} on 2 July 2013 (UT dates are used throughout this paper). 
Analysis of the \textit{Fermi} Gamma-ray Burst Monitor (GBM) light curve showed a burst duration of $T_{90}\approx59$~s (50--300 keV; \citealt{gcn14972}).
The Palomar Transient Factory (PTF) identified an optical afterglow at coordinates $\alpha$ = 14$^{\rm h}$29$^{\rm m}$14.78$^{\rm s}$, $\delta$ = +15$^{\circ}$46$'$26.4$''$ (J2000.0)
at a distance of 3.8$^{\circ}$ from the center of the \textit{Fermi}-GBM trigger 394416326 localization \citep{gcn14967}. 
Follow-up observations of the optical transient designated 
as iPTF13bxl by \citet{gcn14967} yielded an $r'=18.7$ mag estimate at 
4:10 on 2013 July 3. These authors also noted the existence of a 
nearby faint $r=23.01$ mag source SDSS J142914.75+154626.0 separated by 0.6$''$ from the optical transient, 
and a bright red ($r \approx 18$ mag; $g-r \approx 0.9$ mag) galaxy SDSS J142914.57+154619.3, at a separation of 7.6$''$.

\citet{gcn14980} reported optical and infrared measurements of the optical transient with the RATIR telescope,
while CARMA \citep{gcn14979} 
observations detected the transient at millimeter wavelengths. 
RATIR monitoring showed a flattening of the afterglow light curve on 6 and 7 July 2013, five days 
after the \textit{Fermi} detection \citep{gcn14993}. A spectrum taken by \citet{gcn14994} on 8 July 2013 of the optical afterglow shows evidence for broad features, and  \citet{gcn14998} and \citet{gcn15000} obtained additional spectra that suggested that the coincident Type Ic-BL SN was similar to SN 1998bw and SN 2006aj. 

\citet{gcn14985} reported weak H$\alpha$ and [\ionpat{O}{iii}] 
emission lines superimposed on the spectrum of the optical afterglow indicative of
$z=0.145$. A spectrum taken by \citet{gcn14983}
of the bright red galaxy offset by 7.6$''$ from the optical transient 
 showed weak H$\alpha$ and [\ionpat{O}{ii}] emission lines, but 
these authors did not detect emission lines at the explosion site.

\section{Data and Analysis}
\label{sec:data}
The  2.5~m telescope at Apache Point, New Mexico was used to perform the Sloan
Digital Sky Survey (SDSS)
imaging and spectroscopic surveys. The imaging component, consisting of 53.9 s integrations through the SDSS  \textit{ugriz} filter set,  spans 14555 square degrees through Data Release 9 (DR 9; \citealt{ahnalexandroff12}), and 
has an average magnitude limit (95\% point-source completeness) of $r \approx 22.2$ mag.
Each SDSS image has a $13.5' \times 9.9'$ format recorded by a $2048 \times 1498$ pixel array.  
We correct all galaxy magnitudes for extinction by the Galactic dust along the line of sight using the 
\citet{schlaflyfinkbeinerSFD11} recalibration of the \citet{schlegel98} map. 

With the {\it LePhare} code (\citealt{arn99}), we fit SDSS \textit{ugriz} galaxy magnitudes with spectral energy distributions (SEDs) from the PEGASE2 (\citeyear{fi99}) stellar population synthesis models. 
The SEDs include both nebular and stellar light and track metal enrichment and metal-dependent extinction through 69 timesteps for each of the size models. We use the initial mass function (IMF) of \citet{rana92} with a 5\% fraction of
close binaries and a mass range of 0.09 to 120~M$_{\sun}$. 

We acquired two 600~s Keck/DEIMOS \citep{faberdeimos03} spectra of the optical afterglow and adjacent faint source
with the 600 l mm$^{-1}$ grating and 1$''$
slit (full width at half-maximum intensity [FWHM] resolution $\sim3.5$~\AA) in clear conditions on 11 July 2013. 
The spectra were taken with a slit position angle (PA) of 50.4$^{\circ}$ which allowed it to intersect
both the optical transient and the faint adjacent SDSS source that we identify as the dwarf LGRB host galaxy.
The data were taken at an airmass of 1.2, and the seeing was $\sim 1''$ during the observations. 

Immediately following the above pair of integrations,
we obtained a 60~s DEIMOS spectrum with the slit positioned on the center of the bright red galaxy. 
The purpose of this integration, also taken with a PA of 50.4$^{\circ}$, was to 
be able to compare the line-of-sight velocities of the faint source and the bright red galaxy.  

\begin{figure*}[htp!]
\centering
\subfigure{\includegraphics[angle=0,width=5.5in]{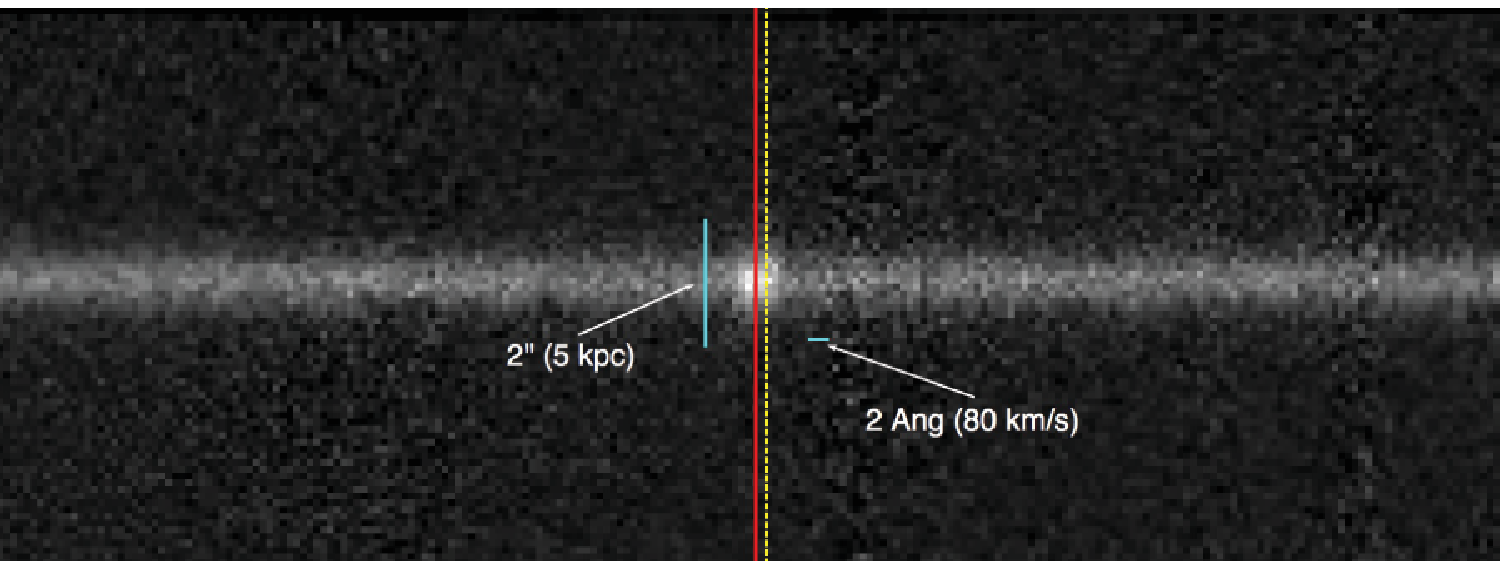}}
\subfigure{\includegraphics[angle=0,width=5.5in]{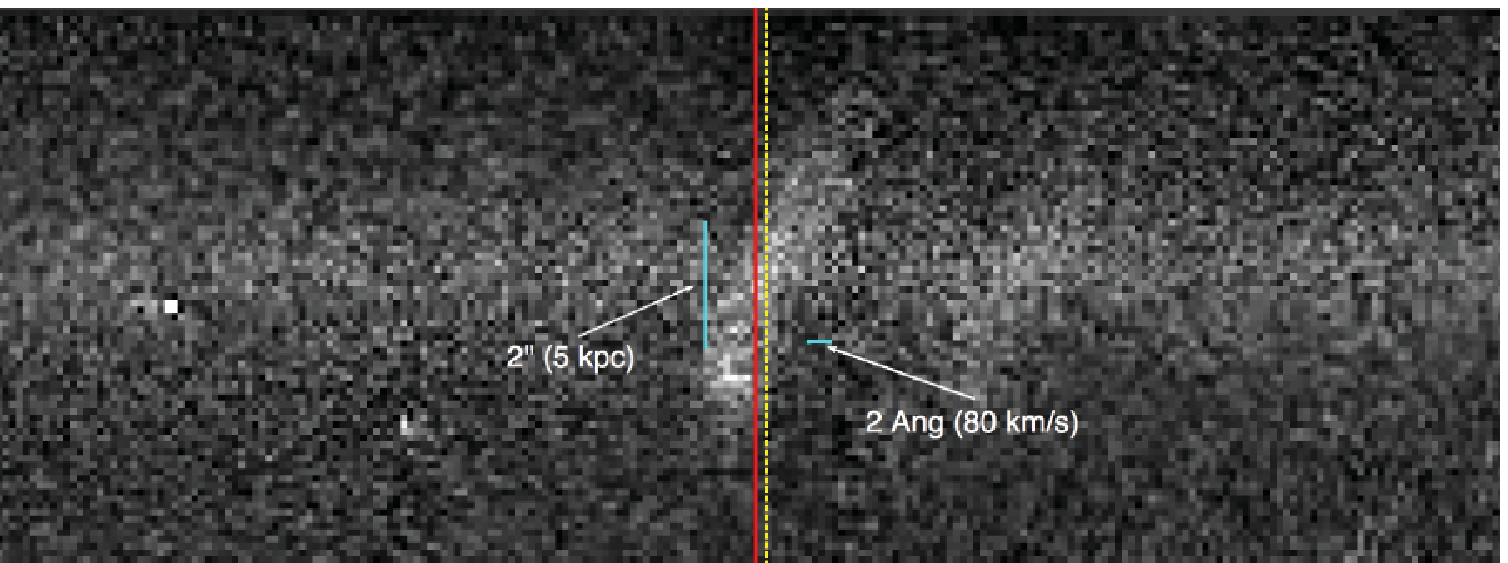}}
\caption{Background- and sky-subtracted Keck spectra of the optical afterglow of GRB 130702A and faint adjacent source S1 (upper panel), and of the bright red galaxy S2 (lower panel). The abscissa corresponds to the dispersion axis, while the ordinate is the spatial axis along the 1$''$ slit.
Each panel shows the trace and superimposed H$\alpha$ emission line (as well as [\ionpat{N}{ii}] in the bottom panel). 
The red solid line marks the \halphasone~\AA\ central wavelength we measure for the H$\alpha$ emission line of the dwarf host galaxy S1, and the yellow dashed line marks the \halphastwo~\AA\ wavelength of the emission line of the bright red galaxy S2. 
The upper panel demonstrates that the H$\alpha$ emission from S1 has the same position along the spatial axis as the continuum light from the optical afterglow, suggesting an association between the LGRB and S1. 
The wavelength profile of the emission lines in the lower panel reflects the rotation of the galaxy's disk. 
}
\label{fig:SONE}
\end{figure*}

All spectra were reduced using IRAF routines and the two-dimensional subtraction techniques developed by \citet{kelson03}. The spectra were flux calibrated using the BD+26$^\circ$2606 spectrophotometric standard star \citep{okegunn83},
and corrections were made for telluric atmospheric absorption features. 
We extracted line strengths after flux calibration 
by fitting a Gaussian to each line profile. The uncertainty of each line-flux estimate was computed using the method described by \citet{perezmonterodiaz03}.
Figure~\ref{fig:SONE} displays image sections of the background- and sky-subtracted spectra centered around the
H$\alpha$ emission lines.


\begin{figure*}[htp!]
\centering
\subfigure{\includegraphics[angle=0,width=3.5in]{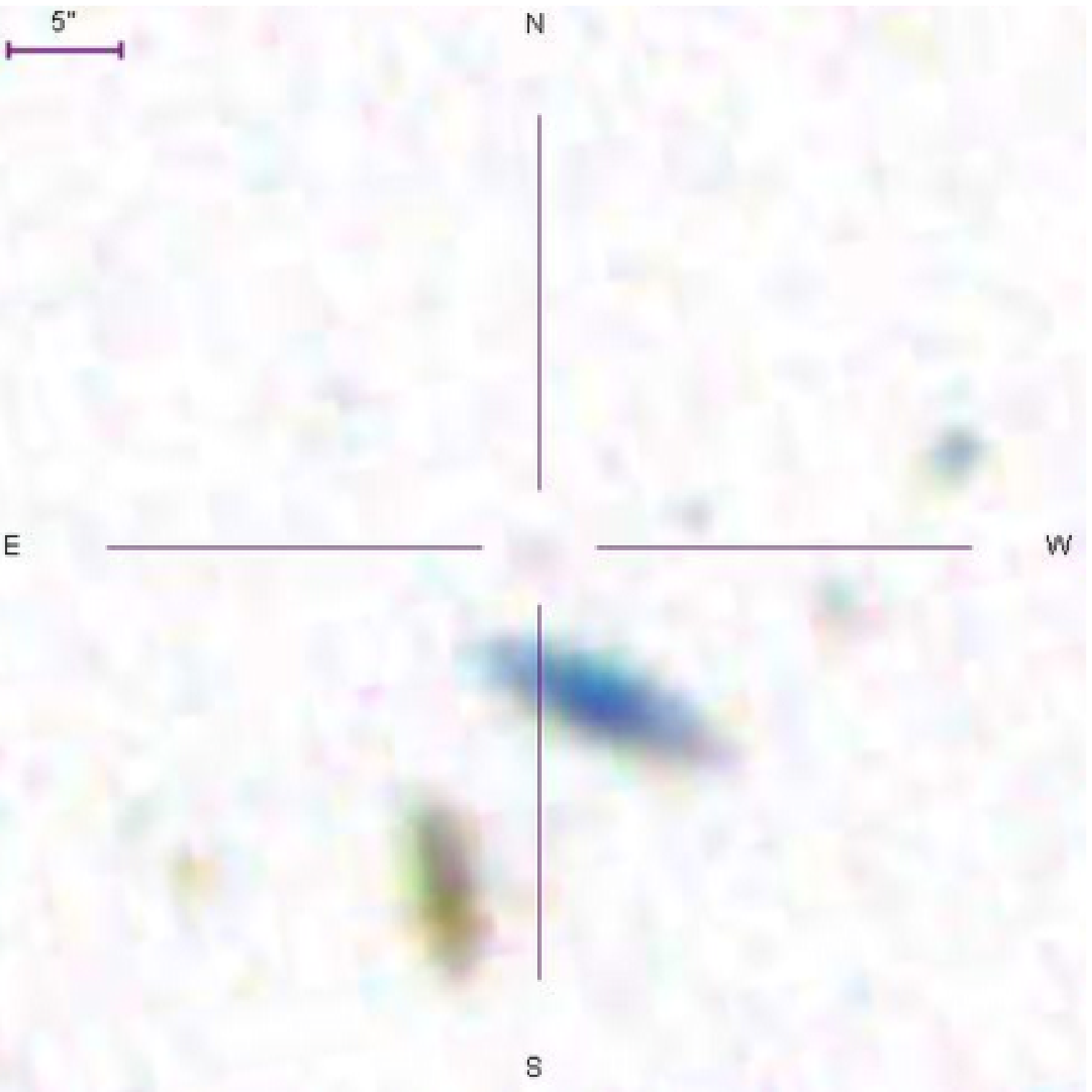}}
\subfigure{\includegraphics[angle=0,width=3.5in]{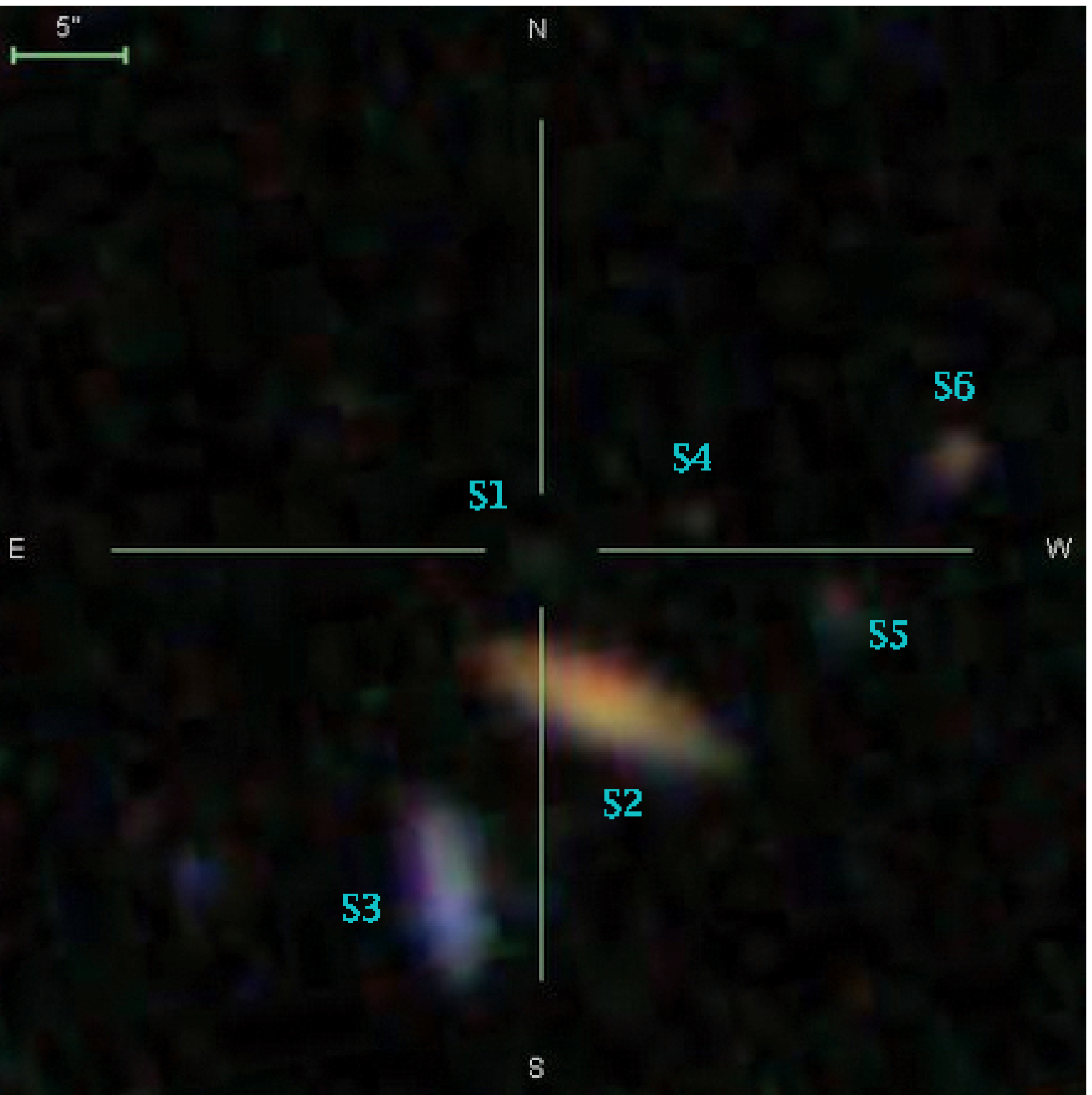}}
\caption{SDSS color composite (left panel) and colormap-inverted (right panel) images of the field surrounding GRB 130702A. 
The position of the LGRB optical afterglow is marked with crosshairs. The center of a faint $r=23.01$~mag 
source S1 is $\sim0.6''$ (projected $\sim1.5$~kpc) from the GRB position, while a 
bright $r=18.14$ mag  galaxy S2 is 7.6$''$ (projected $\sim19.1$ kpc) from the GRB.
The bright blue galaxy S3 ($r=19.02$~mag) has an offset of 15.8$''$ (37.8~kpc) and a 
photometric redshift $0.19\pm0.11$.
The source S4 ($r=22.91$ mag) west of PTF13bxl with offset 7.5$''$ (18.4~kpc in projection) is
unresolved in SDSS images and has no available photometric redshift.
The red source S5 ($r=21.42$ mag) west of PTF13bxl with offset 14.4$''$ (35.6~kpc in projection) has a photometric redshift $0.17\pm0.10$, also consistent with the S1 and S2 redshifts. The red galaxy S6 farther to the north and west with offset 19.1$''$ has a photometric redshift $0.44\pm0.15$, well separated from GRB 130702A.
}
\label{fig:galaxies}
\end{figure*}

\section{Results}
\label{sec:results}

Figure~\ref{fig:galaxies} shows the field surrounding the coordinates of the optical counterpart of GRB 130702A, and 
the label that we assign to each object. 
The faint source S1 closest to the center of the crosshairs is the likely host of the LGRB.
We find evidence that S1 is bound to or interacting with the bright red disk-dominated galaxy (labeled S2) having an offset of 7.6$''$ ($\sim19.1$ kpc in projection) from the
GRB coordinates.
Source S3 is a bright blue galaxy 15.8$''$ ($\sim37.8$~kpc in projection) southeast of the GRB. 
We have no spectroscopic redshift for S3, but its SDSS photometric redshift is $0.19\pm0.11$, consistent with the 0.145 spectroscopic redshift of S1 and S2. 
A large number of faint sources with close proximity to S1 suggest the possibility of local-galaxy interaction.


For the dwarf host galaxy S1, we measure an H$\alpha$ wavelength of \halphasone~\AA~using a Gaussian fit to the line profile. A fit to the broader H$\alpha$ emission profile extracted from an aperture including the central region of the red bright galaxy S2 yields 
an H$\alpha$ wavelength of \halphastwo~\AA.
These are consistent with $z \approx 0.145$ measured by Mulchaey et al. (2013) from 
emission lines superimposed on the afterglow spectrum, and 
are approximately consistent with the redshift expected for the SN Ic-BL associated
with GRB 130702A based on the observer-frame wavelengths of its broad features reported by \citet{gcn14994}.
We use Monte Carlo simulations to compute the uncertainty of the wavelength measurements. 
We estimate that these DEIMOS observations place a 2$\sigma$ upper limit on the 
velocity difference between the S1 and S2 galaxies of $\sim$\,\veloffset. 

From fitting PEGASE2 stellar population synthesis models to the SDSS {\it ugriz} broadband photometry,
we estimate the stellar masses of the faint LGRB host galaxy S1 
(\massfaintpow~M$_{\odot}$), the bright red galaxy S2
(\massbrightpow~M$_{\odot}$), and the blue nearby galaxy S3
($10^{9.5\pm0.08}$ M$_{\odot}$).
This suggests that the massive red galaxy S2 dominates the local gravitational potential, even if the blue
galaxy S3 has close physical proximity. 
Table~\ref{tab:hostparameters} lists the offsets from GRB 130702A, stellar masses, and abundance limits that we estimate for these three galaxies.

The SDSS photometry pipeline performs separate fits of a de Vaucouleurs profile and of an exponential profile to the 
light distribution of each galaxy. 
While holding all parameters from these separate fits constant except the profiles' amplitudes, 
the pipeline next finds the linear combination of the two profiles that has the best agreement 
with flux distribution. 
The logarithmic likelihood of a de Vaucouleurs profile fit to the {\it r}-band light of the bright red galaxy S2 is $-217.3$, exceedingly small, while the logarithmic likelihood of an exponential profile is $-3.4$. 
In the combined fit, the exponential profile receives 98.4\% of the flux.
Using the scale length of the SDSS exponential profile, we compute
that the massive red galaxy S2 has a half-light radius of $\sim$\,\halflight.
Our spectrum of S2 shown in Figure~\ref{fig:SONE} at PA = 50.4$^{\circ}$ 
exhibits evidence for coherent rotation in a disk galaxy with ongoing 
star formation. 


We use the ``expAB\_r'' SDSS pipeline estimate of the ratio $b/a$ (semiminor to semimajor axis lengths) of 
the massive red galaxy S2. If we assume that a very thin disk dominates the galaxy's light (i.e., the disk's thickness does not contribute significantly to the size of the projected semiminor axis),
the best-fitting value of $b/a = 0.22\pm0.06$ corresponds to an inclination angle of $\sim77\pm4^{\circ}$.
The disk thickness or the bulge likely makes a significant contribution to the apparent semiminor axis length, 
so the disk is probably even more inclined. The SDSS exponential fit to the {\it r}-band galaxy flux 
also finds that the angle of the semimajor axis of the galaxy is 67$^{\circ}$ 
counterclockwise north to east. If we make the assumptions that the GRB explosion site is in the plane of the galaxy and that the disk is very thin, then projection of the GRB coordinates onto the plane of the disk yields an offset 
from the galaxy center of $\sim$\,\projoffset. This distance is $\sim$\,\factoroffset~times the half-light radius.
Given the possibility that the disk thickness is significant or that the bulge is substantial, the LGRB offset may be much greater than this value.  

\begin{deluxetable*}{lccc}
\tablecaption{Stellar Masses and Metallicities of Galaxies}
\tablecolumns{4}
\tablehead{\colhead{Name} & \colhead{Separation from Transient}& \colhead{Mass} & \colhead{Metallicity}\\
\colhead{} & \colhead{}& \colhead{log(M/M$_{\odot}$)} & \colhead{12 + log(O/H)}}

\startdata
S1 & 0.6$''$ (1.5 kpc projected on sky) & \massfaint & $\lesssim$8.16 \\
S2 & 7.6$''$ (19.1 kpc projected on sky) & 10.8 $\pm$ 0.1 & \\
S3 & 15.8$''$ (37.8 kpc projected on sky) & 9.5 $\pm$ 0.08\footnote{Only photometric redshift available. We assume that the source has $z=0.145$ for the stellar mass estimate.}& 
\enddata
\tablecomments{Offsets of galaxies from the GRB coordinates, their stellar masses fit from SDSS {\it ugriz} galaxy photometry corrected for Milky Way extinction, and oxygen abundance upper limit for source S1. 
}
\label{tab:hostparameters}
\end{deluxetable*}

The SDSS spectroscopic galaxy survey targeted objects detected by the imaging survey with 5$\sigma$ significance, an extended profile, and an $r'$-band magnitude brighter than 17.77.
Within 15$'$ ($\sim2.3$~Mpc in projection) of the GRB coordinates, five bright galaxies have spectroscopic 
redshifts (0.1427921, 0.1447673, 0.1411451, 0.1418472, 0.1423725) which are similar to the 0.145 redshift of the dwarf LGRB host galaxy and red massive companion.
The standard deviation of the velocities of these bright galaxies is \veldisp~km~s$^{-1}$, 
suggesting that the LGRB host galaxy is a member
of a galaxy group with $\sim2$~Mpc size. 

From the root-mean-square variation of the continuum and the size of the extraction window for dwarf host S1,
we place a 2$\sigma$ upper limit on the [\ionpat{N}{ii}] flux of 
5.4 $\times$ 10$^{-18}$ ergs cm$^{-2}$ s$^{-1}$,
while we find an H$\alpha$ flux of 111 $\times$ 10$^{-18}$ ergs cm$^{-2}$ s$^{-1}$.
These values allow us to place an upper limit on the \citet{pettini04} [\ionpat{N}{ii}]/H$\alpha$ abundance of 12 + log(O/H) $\lesssim$ 8.16 dex.

From the H$\alpha$ = 111 $\pm$ 0.5 $\times$ 10$^{-18}$ ergs cm$^{-2}$ s$^{-1}$
and H$\beta$ = 43 $\pm$ 5.2 $\times$ 10$^{-18}$ ergs cm$^{-2}$ s$^{-1}$ fluxes we measure for the dwarf host galaxy S1,
we estimate a Balmer decrement of H$\alpha$/H$\beta$ = 2.6 $\pm$ 0.3.
This value is consistent with the expectation for Case B recombination (i.e., H$\alpha$/H$\beta$ $=$ 2.86; \citealt{ost89}) and suggests insignificant reddening. We therefore adopt $A_V=0$~mag for the emission-line analysis. 

Applying the \citet{kennicutt98} relation between H$\alpha$ flux and star-formation rate (SFR), we estimate an SFR of 0.05~M$_{\odot}$~yr$^{-1}$ for S1 within the 1$''$ slit aperture. 
For comparison, the SFR of the Large Magellanic Cloud is $\sim0.2$~M$_{\odot}$~yr$^{-1}$ \citep{harriszaritsky09},
while the underlying SFR of the Small Magellanic Cloud, excepting intermittent episodes of stronger
star formation, has been $\sim0.1$~M$_{\odot}$~yr$^{-1}$ \citep{harriszaritsky04}.
The \massfaintpow~M$_{\odot}$ stellar mass of S1 is in the 26th percentile of the 
host masses of 34 LGRBs measured by \citet{svensson10}.
The galaxy SFR is in the 6th percentile of the SFRs of the LGRB host sample, although 
the \citet{svensson10} estimates rely instead on the rest-frame $U$-band luminosity which may 
have a systematic offset from the \citet{kennicutt98} calibration of H$\alpha$ flux. 

\section{Discussion and Conclusions}
\label{sec:discussion}

We have found evidence that the host of GRB 130702A is a dwarf galaxy that is a satellite of a nearby massive, metal-rich galaxy.
Modeling of the two galaxies' photometry suggests that the dwarf galaxy has a stellar mass 
$\sim1$\% that of the red, bright galaxy.
If we assume that the massive red galaxy consists of a very thin exponential disk, and we project the LGRB coordinates onto the disk's inclined plane, 
the transient has an offset from the galaxy center of $\sim$\,\projoffset, or $\sim$\,\factoroffset~times the half-light radius. 
A thick galaxy disk or a significant bulge would imply a greater disk inclination angle and a more remote explosion site.
A large central offset suggests that the faint source is a dwarf galaxy and
not a part of the massive galaxy's disk. 

Comparison between the spectroscopic redshifts of the dwarf and the massive red galaxy shows that 
the velocity offset between the two galaxies is less than $\sim$\,\veloffset, 
which is small in comparison to the $\sim$\,\veldisp~km s$^{-1}$ velocity dispersion 
measured among five bright galaxies within $\sim 2$~Mpc (in projection) of the GRB position having SDSS spectroscopic redshifts. 

We find that the massive red primary galaxy likely dominates the local gravitational potential. 
A blue galaxy is the only bright source that is close in projection, but it 
would have only $\sim10$\% the stellar mass of the red galaxy if it has $z = 0.145$. 
While these observations offer strong evidence that the dwarf host galaxy is a
orbiting satellite of (or merging with) the massive red galaxy, we cannot exclude the possibility of a 
random but improbable coincidence of both their spatial positions and redshifts.

Inspection of images of the field shows a large number of nearby 
faint sources, many of whose photometric redshifts are 
consistent with the GRB spectroscopic redshift. 
A densely populated environment suggests that group members 
experience local interactions.  

Early comparisons between the host environments of nearby LGRBs and star-forming 
galaxies suggested that LGRBs strongly prefer metal-poor
galaxies \citep{stanek06, mod08}, but more recent discoveries and analyses have
found growing numbers of LGRBs that may have metal-rich host galaxies. 
\citet{prochaska07}, for example, used damped Lyman-$\alpha$ (DLA) systems to study the properties of the 
gas along the line of sight to the explosions at the same redshift as the GRB. 
They found evidence for metallicities of the gas from 0.01$Z_{\odot}$ to $Z_{\odot}$. 
\citet{lev10highmet} found that the dark burst GRB 020819B was spatially coincident with a
high-metallicity $z=0.41$ galaxy. 
Recently, \citet{savagliorau12} inferred supersolar 
metallicities from zinc and sulfur absorption features for two $z \approx 3.67$ galaxies
found in the afterglow spectrum of an LGRB.

This analysis of the host galaxy of GRB 130702A provides the first example of 
an LGRB that exploded in a dwarf, metal-poor satellite of a massive, metal-rich galaxy. 
In this case, the dwarf galaxy's projection onto the sky from our perspective on Earth placed it away from the primary galaxy. For other examples of LGRBs that explode in dwarf satellite galaxies, however, 
the dwarf galaxy may be superimposed on a metal-rich galaxy.
This suggests the need to obtain a spectrum at the explosion site of the LGRB to 
attempt to detect the nebular emission of metal-poor gas, as well as high-resolution
follow-up imaging. 

Comparison between the host galaxies of low-redshift SNe~Ic-BL without 
detected GRB emission and normal SNe~Ic whose
features are not broad finds that the SNe~Ic-BL erupt in comparatively metal-poor galaxies (\citealt{kelkir12}; \citealt{sanderssoderberg12}; see also \citealt{arcavi10}). 
This pattern at low redshifts where host galaxies are more accessible 
suggests that low progenitor metallicity may be important to generate
explosions with high ejecta velocities. 
When possible, careful investigation of the immediate environments of LGRBs associated with 
massive galaxies should help to establish how important  
metallicity is to the production of gamma-ray emission.

\acknowledgements
We are very grateful to Brad Cenko for his help in scheduling the observations and reducing the DEIMOS spectra. 
We also thank Maryam Modjaz for her useful instructions for measuring emission-line fluxes,
as well as Daniel Kocevski for discussions about the host-galaxy properties of LGRBs. 
A.V.F.'s group at UC Berkeley has received generous financial assistance
from the Christopher R. Redlich Fund, the TABASGO Foundation, NSF grant
AST-1211916, and NASA/{\it Swift} grant NNX12AD73G.
Some of the data presented herein were obtained at the W. M. Keck
Observatory, which is operated as a scientific partnership among the
California Institute of Technology, the University of California, and
NASA; the observatory was made possible by the generous financial
support of the W. M. Keck Foundation.

\end{document}